\documentclass[%
reprint,
amsmath,amssymb,
aps,
prl,
]{revtex4-2}

\usepackage[all]{background} 
\usepackage{url}
\SetBgContents{Accepted by Physical Review Letters}
\SetBgColor{gray}
\SetBgPosition{4.1,-12}
\SetBgOpacity{0.75}
\SetBgAngle{0.0}
\SetBgScale{2.0}

\usepackage{tikz,xcolor,hyperref}
\definecolor{lime}{HTML}{A6CE39}
\DeclareRobustCommand{\orcidicon}{%
	\begin{tikzpicture}
	\draw[lime, fill=lime] (0,0) 
	circle [radius=0.16] 
	node[white] {{\fontfamily{qag}\selectfont \tiny ID}};
	\draw[white, fill=white] (-0.0625,0.095) 
	circle [radius=0.007];
	\end{tikzpicture}
	\hspace{-2mm}
}
\foreach \x in {A, ..., Z}{%
	\expandafter\xdef\csname orcid\x\endcsname{\noexpand\href{https://orcid.org/\csname orcidauthor\x\endcsname}{\noexpand\orcidicon}}
}

\usepackage{graphicx}
\usepackage{dcolumn}
\usepackage{bm}

\newcommand{\bsub}{\begin{subequations}}
	\newcommand{\esub}{\end{subequations}}

\newcommand{\ep}{\epsilon}

\newcommand{\x}{\times}

\newcommand{\beq}{\begin{equation}}
\newcommand{\eeq}{\end{equation}}
\newcommand{\bsubeq}{\begin{subequations}}
	\newcommand{\esubeq}{\end{subequations}}
\newcommand{\beqn}{\begin{eqnarray}}
\newcommand{\eeqn}{\end{eqnarray}}
\newcommand{\fr}{\frac}
\newcommand{\lb}{\label}
\newcommand{\er}{\eqref}

\usepackage{hyperref}
\hypersetup{colorlinks,allcolors=blue}
\usepackage{amsmath}
\usepackage{multirow}
\usepackage{amsfonts,amssymb}
\usepackage{graphicx}
\usepackage{subfigure}
\usepackage[rel]{overpic}

\graphicspath{{figures/}}

\begin{document}
	
	\preprint{APS/123-QED}
	
	\title{Universal Wind Profile for Conventionally Neutral Atmospheric Boundary Layers}
	
	\author{Luoqin Liu\orcidA{}}
	\email{luoqin.liu@utwente.nl}
	
	\author{Srinidhi N. Gadde\orcidB{}}
	
	\author{Richard J. A. M. Stevens\orcidC{}}
	\email{r.j.a.m.stevens@utwente.nl}
	
	\affiliation{Physics of Fluids Group, Max Planck Center Twente for Complex Fluid Dynamics, University of Twente, 7500 AE Enschede, The Netherlands}%
	
	\date{\today}
	
	\begin{abstract}
		Conventionally neutral atmospheric boundary layers (CNBLs), which are characterized with zero surface potential temperature flux and capped by an inversion of potential temperature, are frequently encountered in nature. Therefore, predicting the wind speed profiles of CNBLs is relevant for weather forecasting, climate modeling, and wind energy applications. However, previous attempts to predict the velocity profiles in CNBLs have had limited success due to the complicated interplay between buoyancy, shear, and Coriolis effects. Here, we utilize ideas from the classical Monin-Obukhov similarity theory in combination with a local scaling hypothesis to derive an analytic expression for the stability correction function $\psi = -c_\psi (z/L)^{1/2}$, where $c_\psi = 4.2$ is an empirical constant, $z$ is the height above ground, and $L$ is the local Obukhov length based on potential temperature flux at that height, for CNBLs. An analytic expression for this flux is also derived using dimensional analysis and a perturbation method approach. We find that the derived profile agrees excellently with the velocity profile in the entire boundary layer obtained from high-fidelity large eddy simulations of typical CNBLs.
	\end{abstract}
	
	\maketitle
	
	\textit{Introduction}.
	For well over a century, wall-bounded turbulent flows have been studied extensively \citep{smi11}. A focus area is the derivation and characterization of the mean velocity profile. In 1925, Prandtl \citep{pra25} recognized that the velocity profile in the inertial sub-layer is approximately logarithmic based on his mixing length hypothesis. In the 1930s, von K\'arm\'an \citep{kar30,kar31} derived the logarithmic law of the wall analytically using dimensional analysis. In 1956, \citet{col56} showed using measurement data that the velocity profile can be described more accurately via the sum of the logarithmic law and a wake function. Since then, the law of the wall has been the pillar of the description of wall-bounded turbulence \citep{ten72, pop00, dav04}. Recently, the universality of the law of the wall has been supported by theoretical and experimental studies \citep{mar13, luc17, sam18}.
	
	The dynamics in atmospheric boundary layers, where most human activity and biological processes occur, are much more complicated as turbulence is generated by shear stress and buoyancy \citep{kat11}, while the Coriolis force creates a wind veer \citep{how20}. In 1954, \citet{mon54} introduced a stability correction function $\psi$ to account for deviations to the logarithmic wind speed profile caused by thermal stratification. Based on the Buckingham $\Pi$ theorem, Monin and Obukhov concluded that $\psi$ is only a function of the atmospheric stability parameter $z/L_w$, where $z$ is the vertical height above the ground and $L_w$ is the Obukhov length based on the surface potential temperature flux. The universality of this well known Monin-Obukhov similarity theory (MOST) \citep{obu46, mon54, mon71} has been established in the surface layers of stable and convective atmospheric boundary layers in many field experiments \citep{fok06}, as well as large eddy simulations (LES) \citep{kha97}. Therefore, the MOST is nowadays regarded as the starting point of modern micro-meteorology \citep{fok06}.
	
	Conventionally neutral atmospheric boundary layers (CNBLs) are also frequently observed and are often considered in fundamental studies \citep{bro82, gra86, tje93, zil02, zil05, zil07, all15, all17}. In contrast to stable and convective atmospheric boundary layers, CNBLs are characterized with zero surface potential temperature flux and capped by an inversion of potential temperature. However, the classical MOST is not applicable to CNBLs, because the surface potential temperature flux is zero, due to which $L_w$ is no longer a relevant scale \citep{kha97}. Many studies \citep{zil02b, esa04, zil05, gry07, kel10, abk13, opt14, kel16, jia18, kel19} have tried to predict the velocity profile in CNBLs, but so far with limited success since none of them have considered the effect of potential temperature flux near the capping inversion layer. Therefore, the logarithmic law of the wall without stability correction is still commonly used to predict the wind speed profile in CNBLs \citep{ros35, bla62, bla68, ten73, hor12}. However, an analytical description of the velocity profiles in CNBLs is of great fundamental interest and relevant for meteorological applications \citep{kat11, kel19}.
	
	\textit{Theory}. In this work, we derive the potential temperature flux profile using dimensional analysis and a perturbation method approach. To account for the deviation to the logarithmic wind speed profile we use ideas from the MOST in combination with a local similarity hypothesis. Therefore, we introduce a stability correction function $ \psi $ that depends only on the local stability parameter $z/L $, where $ L $ is the local Obukhov length based on the local potential temperature flux. The canonical shape of $ \psi $ is determined by asymptotic analysis. The derived universality profiles for the potential temperature flux and the wind speed profiles are confirmed by the excellent agreements with the results of high-fidelity LES.
	
	Based on the dimensional analysis (a derivation can be found in section III of the Supplementary Material \citep{liu20e2}), the potential temperature flux can be written as
	\beq \lb{eq.q}
	\fr{ \beta z_0 q}{ u_*^3 } = - \Pi \left(Ro, Zi, \xi \right) = - {Ro}^r {Zi}^s \Pi_1 (\xi).
	\eeq 
	Here $q$ is the potential temperature flux, $u_*$ is the friction velocity, $\beta$ is the buoyancy parameter, $\Pi$ and $\Pi_1$ are dimensionless functions, $Ro=u_*/(|f|z_0)$ is the Rossby number with $f$ the Coriolis parameter and $z_0$ the roughness height, $Zi=N/|f|$ is the Zilitinkevich number \citep{esa04} with $N$ the free-atmosphere Brunt–V\"ais\"al\"a frequency, and $r$ and $s$ are the power exponents for $Ro$ and $Zi$, respectively. The dimensionless parameter $\xi = z/h'$, where $h'= {h}/{(1-0.05^{2/3})}$ and $h$ is the boundary layer height at which the total momentum flux reaches $5\%$ of the surface value. This conventional definition underestimates the actual boundary layer height. The functional form of $h'$ follows from the fact that the dimensionless total momentum flux follows a power law with exponent $3/2$ \cite{nie84}. We note that the first equality in Eq.~\er{eq.q} is unconditionally valid, although the corresponding functional form is hard to determine analytically. The second equality is valid under the presumed power-law dependence of $\Pi$ on the two independent dimensionless parameters $Ro$ and $Zi$. The values of $s$ and $r$ will be determined later from our high-fidelity LES data.

	We emphasize that other definitions for the boundary layer height, which are based on, for example, the vertical wind speed or potential temperature profiles are also commonly used \citep{abk13, all15, kel19}. In particular, the boundary layer height $h_t$ is defined as the height at which the potential temperature flux reaches its minimum value. Previous studies \citep{mau07, ped14, all15, kel19, ber20} showed that the potential temperature flux in CNBLs decreases linearly from zero at the surface to a minimum value at $z=h_t$, and then increases to zero for $z\ge h'$. As explained in section IV of the Supplementary Material \citep{liu20e2}, the ratio $h_t/h' \equiv 1-2 \epsilon$ is a function of $Zi$. However, as we will see later the dependence of $\epsilon$ on $Zi$ is limited over the parameter regime under consideration. Clearly, $\ep\ll 1$ represents the half capping layer thickness normalized by the height $h'$, where the potential temperature flux recovers steeply to zero. Therefore, we propose the following ordinary differential equation to model the potential temperature flux,
	\beq\lb{eq.pi1}
	-\epsilon \Pi_1'' + \Pi_1' = c_\Pi, \quad \Pi_1(0)=\Pi_1(1)=0.
	\eeq 
	Here $c_\Pi$ is the slope of the dimensionless total potential temperature flux $ -(\beta z_0 q)/ (u_*^3 Zi^s Ro^{r})$ in the surface layers, which can be determined from simulation or measurement data. The solution of $\Pi_1$ reads
	\beqn\lb{eq.pi1-final}
	\Pi_1 = \left\{ 
	\begin{split}
		& c_\Pi \left( \xi - \fr{ e^{\xi/\ep} - 1}{ e^{1/\ep} - 1} \right), & \quad \xi \le 1, \\
		& 0, & \quad \xi >1. 
	\end{split}
	\right.
	\eeqn
	Note that Eq.~\er{eq.pi1} and its solution Eq.~\er{eq.pi1-final} is reminiscent of the classical singular perturbation method \citep{dyk75}: The outer solution (close to the wall) is a linear function of $\xi$ and the inner solution (close to the capping inversion layer) is controlled by a small parameter $\ep$.

	In contrast to the classical MOST \citep{mon54} where the normalized wind speed gradient is assumed to be a universal function, we introduce a stability correction function $\psi$ to account for the deviation of the logarithmic profile. Therefore we write the wind speed profile as
	\beq \lb{eq.umag}
	\fr{\kappa U_{\rm mag}}{u_*} = \ln \left( \fr{z}{z_0} \right) - \psi \left( \fr{z}{L} \right),
	\eeq 
	where $\kappa=0.4$ is the von K\'arm\'an constant, $U_{\rm mag}$ is the mean wind speed, and $\psi$ is the stability correction function that depends only on the dimensionless stability parameter $z/L$. According to the local scaling hypothesis \citep{nie84, sor86}, $L$ is defined as the local Obukhov length,
	\beq \lb{eq.L}
	\fr{1}{L} \equiv - \fr{ \kappa \beta q }{ u_*^3 }.
	\eeq 
	It is worth to point out that the dimensionless slope $(\kappa z/u_*) \textrm{d} U_{\rm mag} / \textrm{d} z$ is usually regarded as a universal function of the stability parameter $z/L$ in the stable and convective atmospheric boundary layers \citep{bus71}. However, under the assumption of Eq.~\er{eq.umag}, this slope is no longer a universal function of $z/L$. 
	
	To determine the canonical shape of $\psi$, we assume
	\beq \lb{eq.psi}
	\psi = -c_\psi \left( \fr{z}{L} \right)^{p},
	\eeq 
	where $ p $ is the power exponent to be determined analytically below, and $c_\psi$ is an empirical constant. Recall that very close to the wall (see Eq.~\er{eq.pi1-final})
	\beq \lb{eq.zoverL-1}
	\Pi_1 \rightarrow c_\Pi \xi = c_\Pi \fr{z}{h'} \quad \text{as} \quad \fr{z}{h'} \rightarrow 0.
	\eeq
	Then, from asymptotic analysis \citep{lin88}, we find that
	\beq \lb{eq.zoverL}
	\fr{z}{L} = \fr { \kappa z}{ z_0 }  \Pi \rightarrow  \fr{ c_\Pi \kappa h' }{z_0} Zi^s Ro^r  \left( \fr{z}{h'} \right)^2 \quad \text{as} \quad \fr{z}{h'}  \rightarrow 0.
	\eeq 
	\citet{zil05} showed that in the surface layers of stable, truly neutral, and conventionally neutral atmospheric boundary layers $\psi = - C_u z/L_M $. Here $C_u$ is a dimensionless constant and $L_M$ is the combined turbulent length scale, which in CNBLs can be estimated as $|f| L_M / u_* = (1+C_m^2 Zi^2)^{-1/2}$, where $C_m$ is an empirical constant (\citep{zil05}, see also the Supplementary Material \citep{liu20e2}). To match with the result of \citet{zil05} in the surface layer, we find that $p=1/2$. Clearly, the determination of $p$ is independent of the values of $c_\Pi$, $c_\psi$, $r$, $s$, and $\epsilon$. Thus, the wind speed profile is given by
	\beqn\lb{eq.most}
	\fr{\kappa U_{\rm mag}}{u_*} = \left\{ 
	\begin{split}
		& \ln \left( \fr{z}{z_0} \right) + c_\psi \left( \fr{z}{L} \right)^{1/2}, & \quad \xi \le \xi_0, \\
		& \fr{\kappa G}{u_*}, & \quad \xi >\xi_0.
	\end{split}
	\right.
	\eeqn 
	Here $G$ is the geostrophic wind speed, $\xi_0$ is the highest intersection point of the curves described by the upper and lower expressions in Eq.~\er{eq.most}, $z/L$ is the dimensionless stability parameter predicted by the potential temperature flux model (i.e. Eqs.~\er{eq.q} and \er{eq.pi1-final}), and $c_\psi$ is the empirical constant that can be determined from simulation or measurement data. 
	
	\textit{Validation}. 
	To verify the universality of the wind speed profile for CNBLs, we perform six high-fidelity LES. In the simulations, a CNBL over a flat surface with periodic conditions in horizontal directions is considered. The flow is initialized with uniform geostrophic wind speed and a linear potential temperature profile with a constant gradient \cite{ped14,kel19}. The simulations are performed with an in-house code \citep{bou05, ste14d, nag19, nag20, liu20b, liu20d}, which employs a pseudo-spectral discretization in the horizontal directions and a second-order finite difference method in the vertical direction. We employ the advanced anisotropic minimum dissipation model to parameterize the sub-grid scale shear stress and potential temperature flux \citep{abk17}. The horizontal domain size is more than six times larger than the boundary layer height, and the grid resolution is $288^3$. We ensure that all simulations have reached the quasi-stationary state and the statistics are averaged over one inertial period \citep{col92}. A summary of all simulated cases is presented in Table~\ref{tab.summary}. The simulated Zilitinkevich number $Zi$ and Rossby number $Ro$ range covers the values found in typical CNBLs at mid to high latitudes \citep{hes02-a, zil12}. More details about the numerical method and simulation setup can be found in the Supplementary Material \citep{liu20e2}.

	\begin{table}[!tb]
		\centering
		\caption{Summary of all simulated cases, where the $Zi$ and $Ro$ range covers the values found in typical CNBLs at mid to high latitudes \citep{hes02-a, zil12} }\lb{tab.summary}
		\begin{tabular}{cccccccccc}
			\hline 
			\hline 
			Case no. & A & B & C & D & E & F \\
			\hline 
			$Zi$ & 51.2 & 88.7 & 88.7 & 88.7 & 88.7 & 153.6 \\
			$Ro$ & 2.7E7 & 4.5E4 & 3.7E5 & 3.2E6 & 2.7E7 & 2.7E7 \\
			$\epsilon$ & 0.1186 & 0.1148 & 0.1191 & 0.1213 & 0.1224 & 0.1341 \\
			$c_\Pi$ & 0.0335 & 0.0335 & 0.0330 & 0.0329 & 0.0330 & 0.0336 \\
			\hline
			\hline 
		\end{tabular}
	\end{table}

	\begin{figure} [!tb]
		\centering
		\begin{overpic}[width=0.4\textwidth]{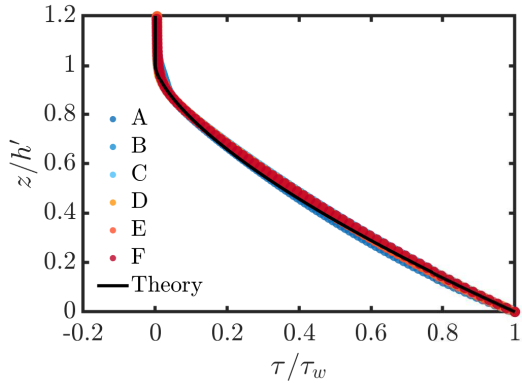} 
		\end{overpic}
		\caption{Vertical profile of dimensionless mean total momentum flux $\tau/\tau_w$. Filled symbols: LES data; solid line: theoretical curve given by $\tau/\tau_w = (1-z/h')^{3/2}$. }
		\label{fig.tau} 
	\end{figure}

	Figure~\ref{fig.tau} shows the vertical profile of the dimensionless mean total momentum flux $\tau/\tau_w$, where $\tau$ is the total momentum flux and $\tau_w$ is its surface value. All different cases in Table~\ref{tab.summary} are shown in the figure (filled symbols). \citet{nie84} analytically determined that the total momentum flux profile in stable atmospheric boundary layers scales as $\tau/\tau_w = (1-z/h)^{3/2}$. In Fig.~\ref{fig.tau} we show that this expression is still valid for CNBLs when we consider the previously introduced boundary layer thickness $h'= {h}/{(1-0.05^{2/3})}$. The finding that the dimensionless momentum flux profiles obtained from all LES collapse to the theoretical curve (see Fig.~\ref{fig.tau}) confirms that $h'$ is the appropriate boundary layer height scale to consider.

	To determine the values of the power indices $r$ and $s$, we take the vertical derivative of Eq.~\er{eq.q}, see details in section III of the Supplementary Material \citep{liu20e2}. Figure~\ref{fig.q-r-s} shows the dimensionless mean potential temperature gradient $\ln \, (- \beta z_0 q'  / u_*^3)$ versus (a) the Rossby number $\ln Ro$ and (b) the Zilitinkevich number $\ln Zi$ in the surface layer, where $q'=\textrm{d} q/\textrm{d} \xi$. The slopes of the curve shown in the figure determine the values of the power exponents $r$ and $s$. In the parameter regime under consideration $r=-1.002 \approx -1$ and $s=1.004 \approx 1$ describe the data very well. We note that the values of $c_\Pi$ can also be determined from the figure and the results are listed in Table~\ref{tab.summary}.

	\begin{figure}[!tb]
		\centering
		\begin{overpic}[width=0.42\textwidth]{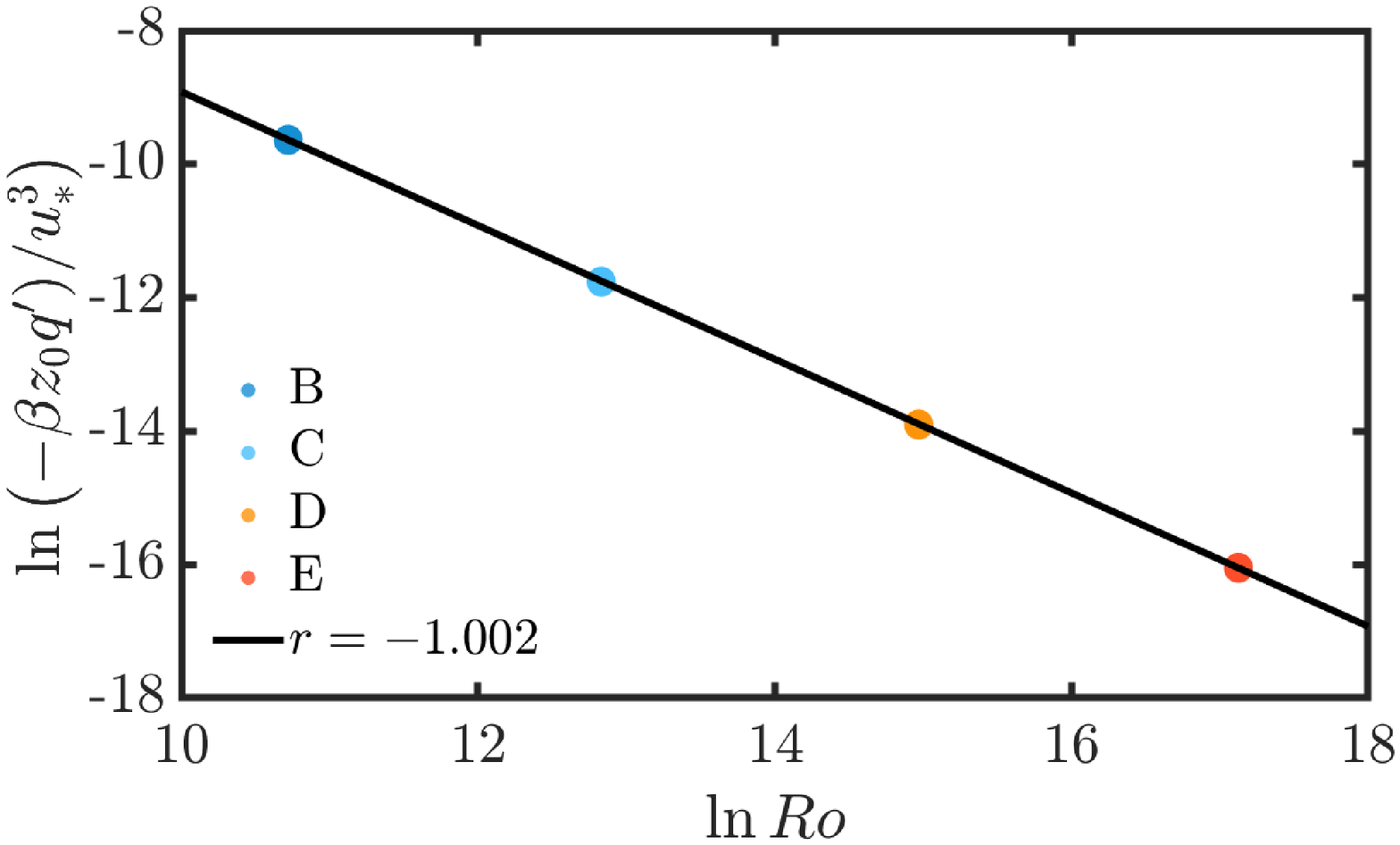} 
			\put(0,53){$(a)$}
		\end{overpic}
		\begin{overpic}[width=0.42\textwidth]{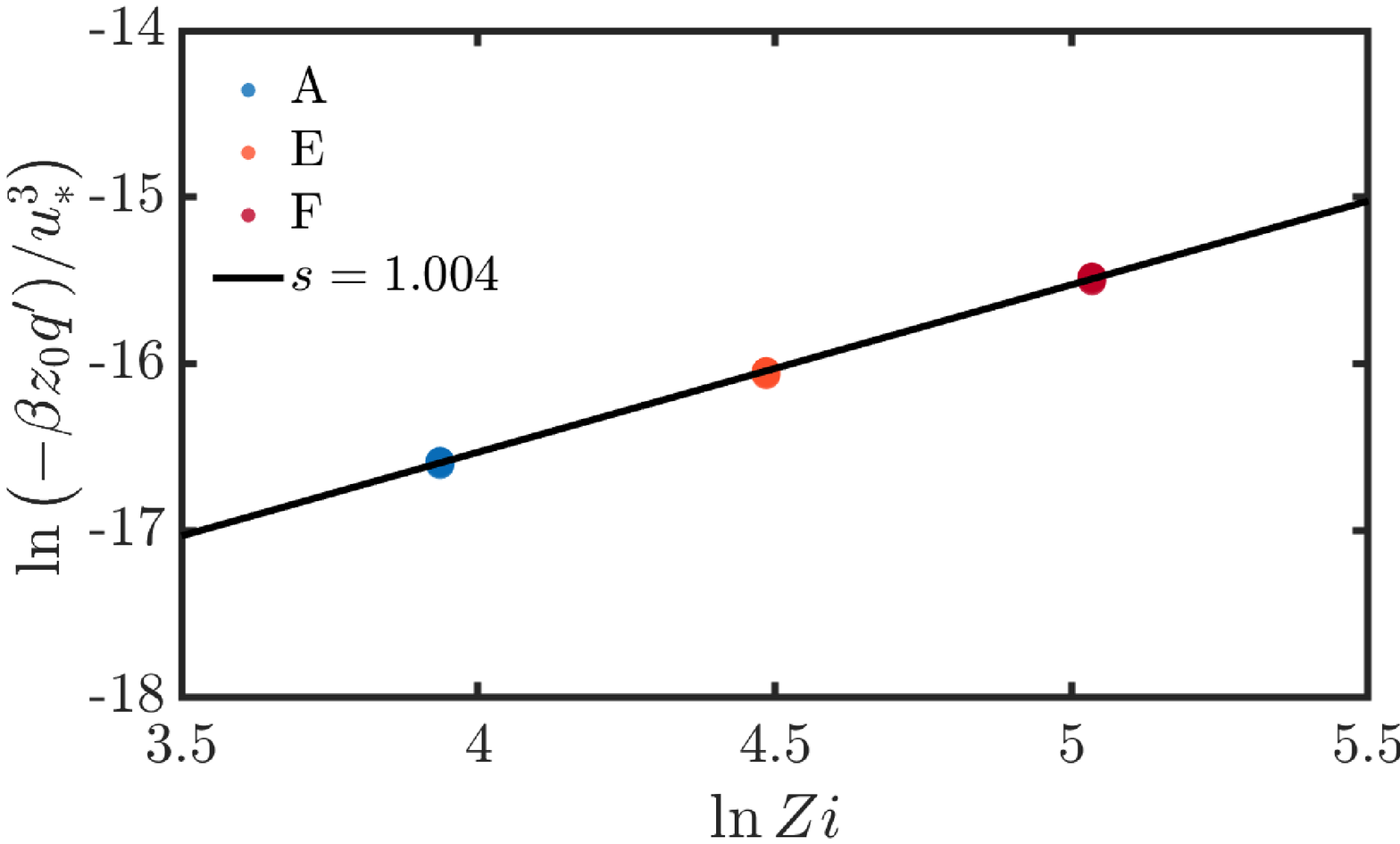} 
			\put(0,53){$(b)$}
		\end{overpic}
		\caption{Dimensionless mean potential temperature gradient in the surface layer versus (a) the Rossby number $Ro$ and (b) the Zilitinkevich number $Zi$, where $q'= \textrm{d} q/ \textrm{d} \xi$. The values of the slope are (a) $r=-1.002 \approx -1$ and (b) $s=1.004 \approx 1$, which are determined by a least-squares fitting procedure. }
		\label{fig.q-r-s} 
	\end{figure}

	\begin{figure}[!thb]
		\centering
		\begin{overpic}[width=0.45\textwidth]{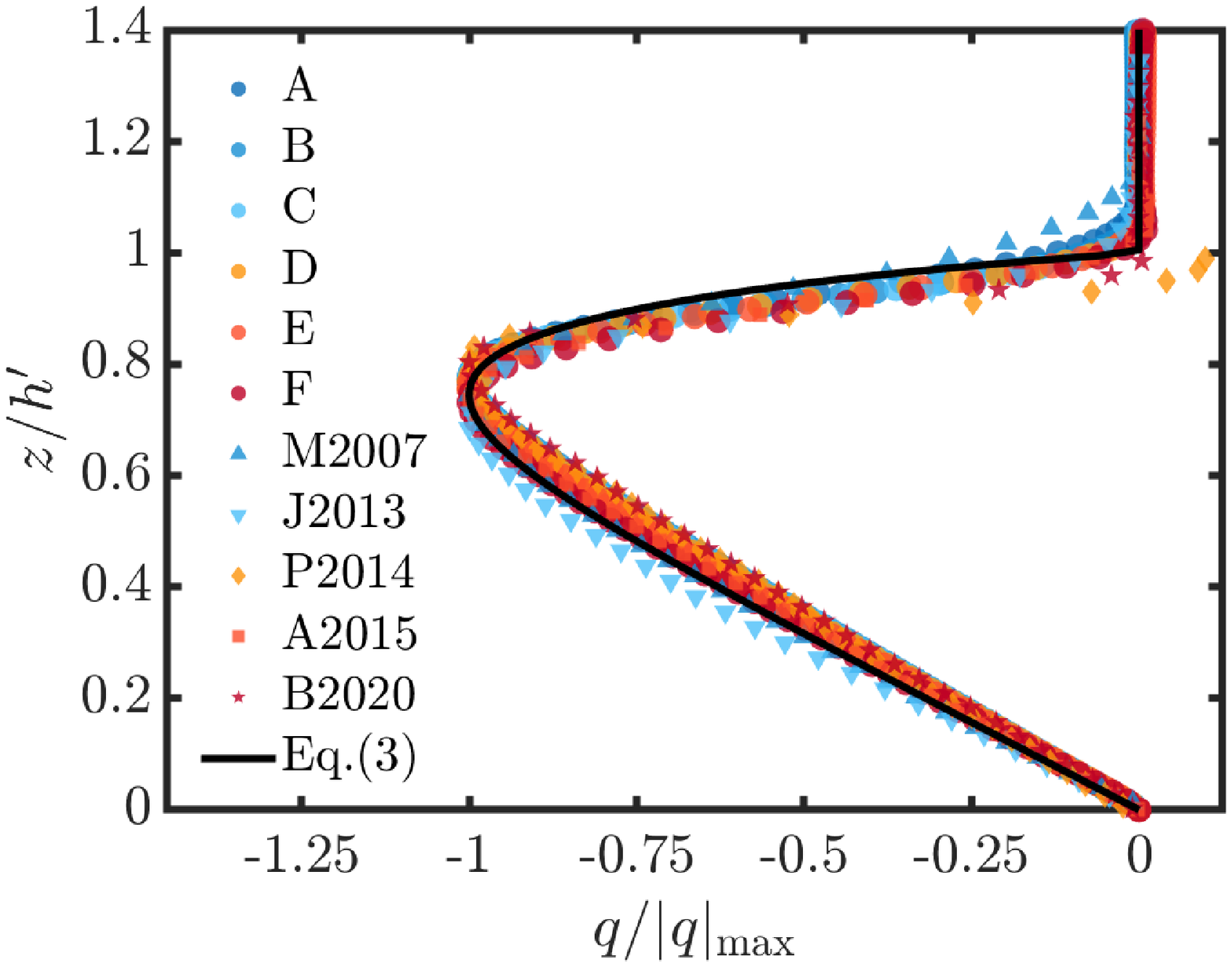} 
		\end{overpic}
		\caption{Vertical profile of dimensionless mean potential temperature flux $q / |q|_{\rm max}$. Filled circles: LES data of the present work; filled triangles: prediction based on a turbulence closure given by \citet{mau07}; filled inverted triangles: DNS data of \citet{jon13}; filled diamonds: LES data of \citet{ped14}; filled squares: LES data of \citet{all15}; filled stars: LES data of \citet{ber20}; solid line: theoretical prediction given by Eq.~\er{eq.pi1-final} with $\ep=0.12$. }
		\label{fig.q} 
	\end{figure}

	Figure~\ref{fig.q} shows the vertical profile of the dimensionless mean potential temperature flux $q / |q|_{\rm max}$, which reduces to almost zero at $z/h' \ge 1$. The potential temperature flux first decreases linearly from zero at the surface to a minimum value at $z= h_t \equiv (1-2\epsilon)h'$, and then increases rapidly to zero in a narrow region ($1-2\epsilon \le z/h' \le 1$) since $\epsilon \ll 1$. The value of $\epsilon$ is expected to depend only on $Zi$ (see the Supplementary Material \citep{liu20e2}). The data in Table~\ref{tab.summary} show that for the parameter range under consideration the variation of $\epsilon$ is limited and therefore we take $\epsilon=0.12$ to describe the data here. Evidently, all LES data of the present work (filled symbols) collapse very well to the introduced theoretical model (solid line), which validates the chosen approach. For comparison, the prediction based on a turbulence closure given by \citet{mau07}, the direct numerical simulations (DNS) data performed by \citet{jon13}, and the LES data taken from \citet{ped14}, \citet{all15}, and \citet{ber20} are also shown in the figure. The overall agreement between the theoretical prediction and the data from previous studies \citep{mau07, jon13, ped14, all15, ber20} is very good, which confirms the universality of the proposed potential temperature flux profile.

\begin{figure}[!tb]
		\centering
		\begin{overpic}[width=0.42\textwidth]{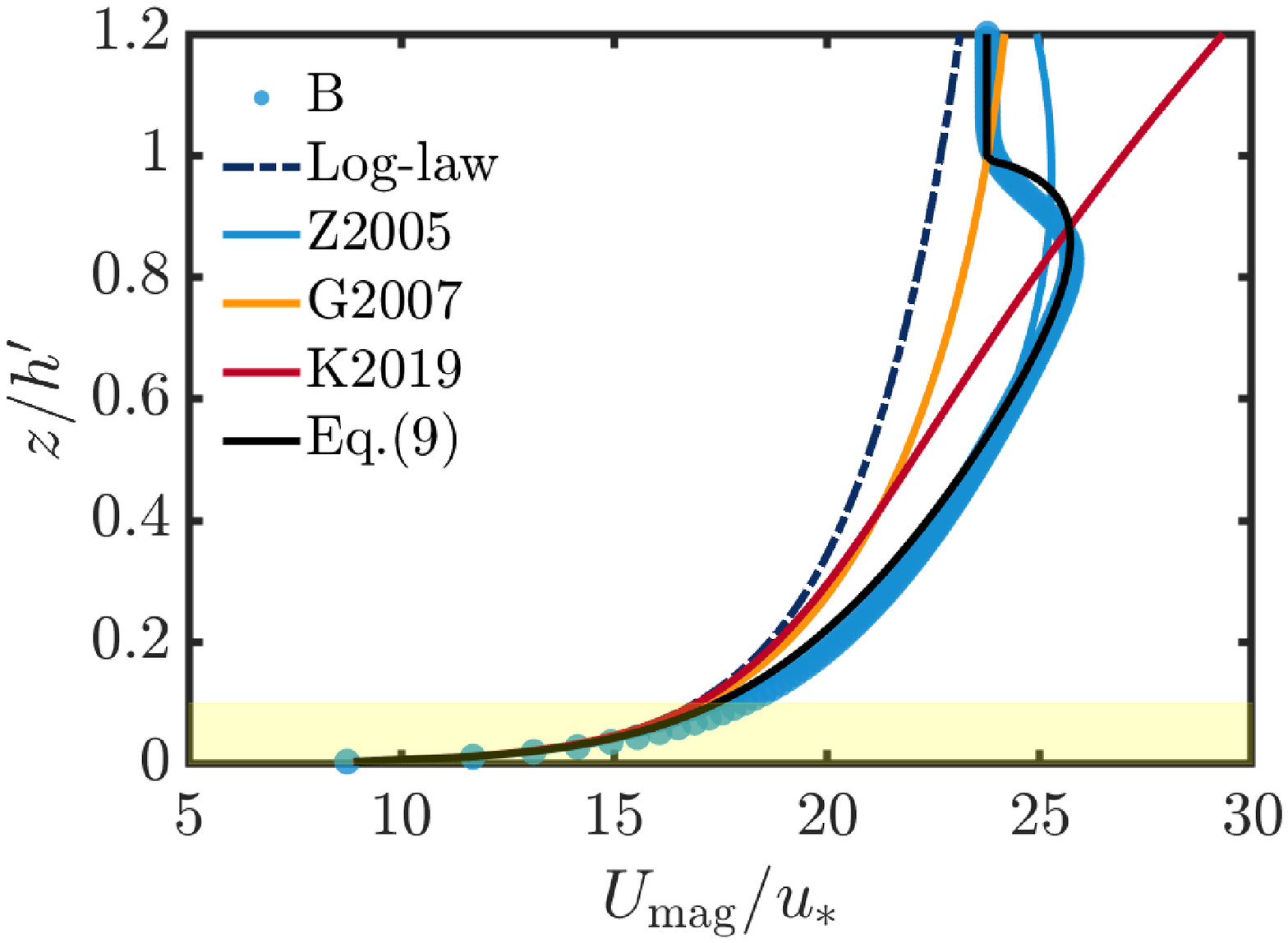} 
			\put(0,54){$(a)$}
		\end{overpic}
		\begin{overpic}[width=0.42\textwidth]{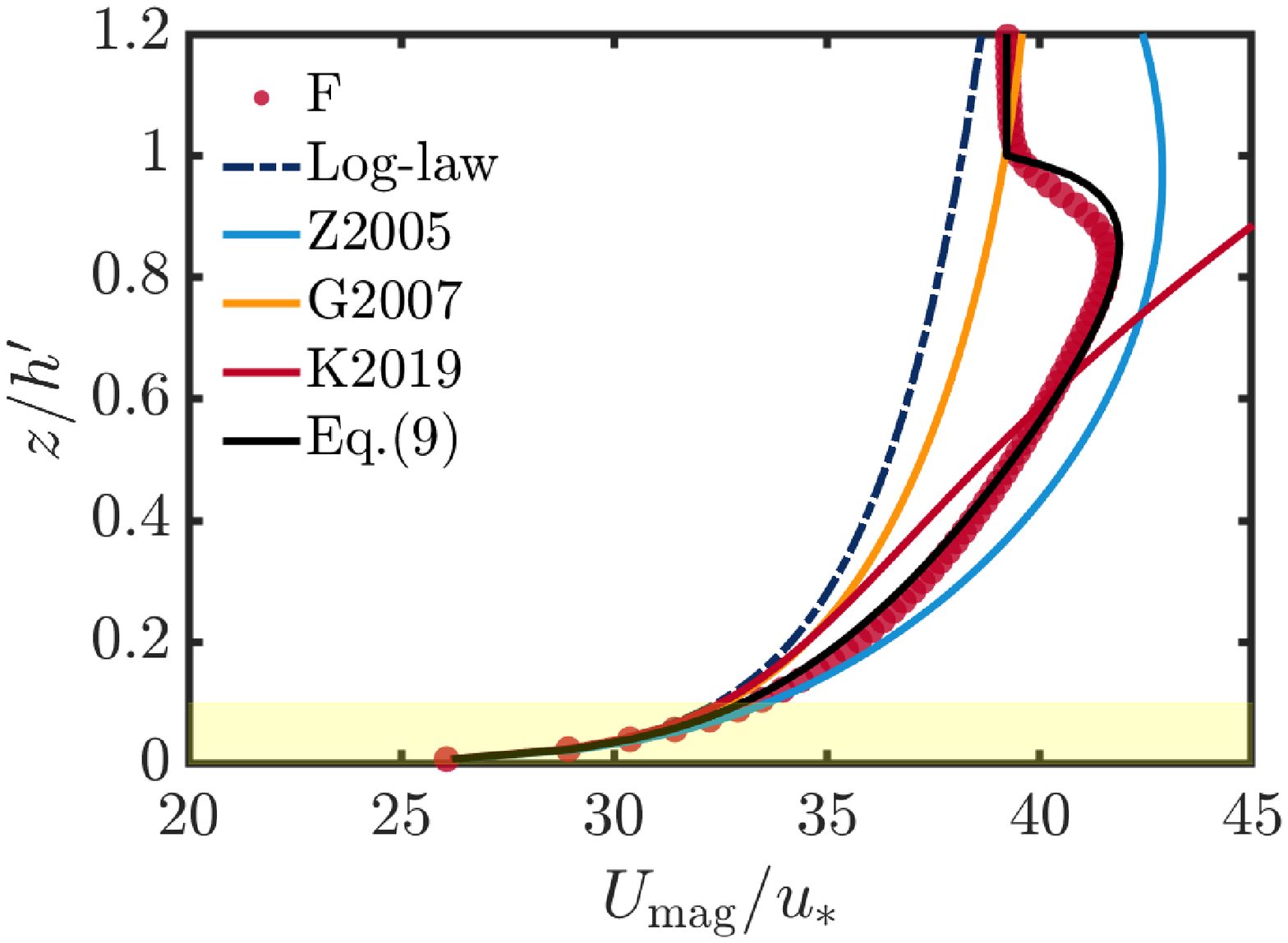} 
			\put(0,66){$(b)$}
		\end{overpic}
		\caption{Vertical profile of mean wind speed for (a) case B and (b) case F. Filled symbols: LES data; dashed line: prediction given by the logarithmic law; blue line: prediction given by \citet{zil05}; yellow line: prediction given by \citet{gry07}; red line: prediction given by \citet{kel19}; black line: prediction given by Eq.~\er{eq.most} with $c_\psi=4.2$. }
		\label{fig.umag} 
	\end{figure}

	Figure~\ref{fig.umag} shows the vertical profile of the dimensionless wind speed for two typical cases, which covers the $ Zi $ and $ Ro $ number range of typical CNBLs at mid to high latitudes \citep{hes02-a, zil12}. The filled symbols are LES data, the dashed line is the theoretical prediction given by the logarithmic law, the blue line is the prediction of \citet{zil05}, the yellow line is the prediction of \citet{gry07}, the red line is the prediction of \citet{kel19}, and the black line is the prediction given by Eq.~\er{eq.most} with $c_\psi=4.2$ where the potential temperature flux profile is modeled by Eq.~\er{eq.pi1-final} with $c_\Pi=0.0332$ and $\ep=0.12$ (see Table~\ref{tab.summary}). The empirical constant $c_\psi$ is determined such that it can predict the wind speed profiles of all cases in Table~\ref{tab.summary} with minimum discrepancies. The figure shows that the logarithmic law only accurately captures the wind speed in the lower 10\% of the boundary layer, also known as the surface layer (shaded region). The theory given by \citet{gry07} focuses on capturing the wind speed at the top of the CNBL, but does not capture the effect of the low-level jet. The theory given by \citet{kel19} is focused on the lower part of the CNBL. The predictions by \citet{zil05} agree well with the LES data in the lower part of the CNBL, but do not capture the low-level jet, which is represented in our approach. In contrast, the agreement between the proposed profile \er{eq.most} and the LES data is nearly perfect in the entire boundary layer and much better than all previous approaches. This excellent agreement confirms the universality of our proposed wind profile \er{eq.most} in the considered parameter range of CNBLs.

	\textit{Summary}. 
	We propose a universal velocity profile for CNBL derived using a local similarity hypothesis combined with ideas from the classical Monin-Obukhov similarity theory. We introduce a stability correction function $ \psi $ to account for the deviation of the logarithmic law. The canonical shape of $\psi$ is determined theoretically as $\psi =-c_\psi (z/L)^{1/2}$, where $c_\psi=4.2$ is an empirical constant determined from simulation data, $z$ is the vertical height above the surface, and $L$ is the local Obukhov length. An analytical expression for the potential temperature flux profile is also derived from dimensional analysis and perturbation method. The universality of the proposed profile \er{eq.most} has been confirmed by its excellent agreement with high-fidelity LES results for $Ro= [4.5\times10^4, 2.7 \times 10^7]$ and $Zi \in [51, 154]$, where the $Zi$ and $Ro$ number range cover the range of values observed in typical CNBLs at mid to high latitudes. 
	Further work is required to assess the applicability of the approach to other parameter regimes.

	\begin{acknowledgments}
		We appreciate very much the valuable comments of the anonymous referees. We acknowledge Drs. K. L. Chong and Y. X. Li for insightful discussion. This work is part of the Shell-NWO/FOM-initiative Computational sciences for energy research of Shell and Chemical Sciences, Earth and Live Sciences, Physical Sciences, FOM and STW, and an STW VIDI grant (No.\ 14868).\ This work was carried out on the national e-infrastructure of SURFsara, a subsidiary of SURF cooperation, the collaborative ICT organization for Dutch education and research. 
	\end{acknowledgments}
	
	\bibliography{literature_windfarms}
	
\end{document}


\preprint{APS/123-QED}
	
	\title{Supplementary Material}
	
	\author{Luoqin Liu\orcidA{}}
	\email{luoqin.liu@utwente.nl}
	
	\author{Srinidhi N. Gadde\orcidB{}}
	
	\author{Richard J. A. M. Stevens\orcidC{}}
	\email{r.j.a.m.stevens@utwente.nl}
	
	\affiliation{Physics of Fluids Group, Max Planck Center Twente for Complex Fluid Dynamics, University of Twente, 7500 AE Enschede, The Netherlands}%
	
	\date{\today}
	
	
	
	\maketitle

	In sections~\ref{sec1} and \ref{sec2}, we provide additional information on the used numerical method and the simulations' setup and validation, respectively. Sections~\ref{sec3} and \ref{sec4} provide a derivation of Eq.~(1) of the manuscript and $\epsilon$ dependence on the Zilitinkevich number, respectively. A summary of previous models for CNBLs is given in section \ref{sec5}.

	\section{Numerical method}\label{sec1}
	
	We perform large eddy simulations (LES) of conventionally neutral atmospheric boundary layers (CNBLs) over a flat surface with homogeneous roughness. We integrate the spatially-filtered Navier-Stokes equations and the filtered transport equation for the potential temperature \citep{alb96, alb99, nag20}: 
	\beqn
	& & \pat_t \widetilde{\pu} + \widetilde{\po} \times \widetilde{\pu} = f \pe_z \times (\pG - \widetilde{\pu}) + \beta (\widetilde \theta - \langle \widetilde \theta \rangle) \pe_z - \na \widetilde {p} - \na \cdot \pt, \quad 
	\na \cdot \widetilde{\pu} = 0, \lb{eq.ns} \\
	& & \pat_t \widetilde{\theta} + \widetilde{\pu} \cdot \na \widetilde{\theta} = - \na \cdot \widetilde {\pq}. \lb{eq.temp}
	\eeqn
	Here, the tilde denotes spatial filtering at a scale of $\Delta=(\Delta x \Delta y \Delta z)^{1/3}$, $\langle \cdot \rangle$ represents horizontal averaging, $\widetilde \pu$ is the velocity, $\widetilde \po = \na \times \widetilde \pu$ is the vorticity, $\widetilde p$ is the modified pressure, $\widetilde \theta$ is the potential temperature, $f$ is the Coriolis parameter, $\beta$ is the buoyancy parameter, $\pG$ is the geostrophic wind velocity, $\pt$ denotes the deviatoric part of the sub-grid scale (SGS) shear stress, and $\pq$ represents the SGS potential temperature flux. Molecular viscosity and diffusivity are neglected as the Reynolds number in the CNBL flow is very high. The SGS shear stress and potential temperature flux are parameterized using the recently developed anisotropic minimum dissipation model \citep{abk17}. 
	
	Our code is an updated version of the one used by \citet{alb99}. The grid points are uniformly distributed, and the computational planes for horizontal and vertical velocities are staggered in the vertical direction. The first vertical velocity grid plane is located at the ground, and the first streamwise and spanwise velocities and potential temperature grid planes are located at half a grid distance away from the ground. We use a second-order finite difference method in the vertical direction, while we use a pseudo-spectral discretization with periodic boundary conditions in the horizontal directions. Time integration is performed using the second-order Adams-Bashforth method. The projection method is used to ensure the divergence-free condition of the velocity field.
	
	At the top boundary we enforce a constant potential temperature gradient, zero vertical velocity, and zero shear stress boundary condition. At the bottom boundary, we enforce a zero potential temperature flux and the classical wall stress formulation \citep{moe84, bou05},
	\beq
	\tau_{xz} = - \left( \frac{\kappa}{\ln z_1/z_0} \right)^2 \left( \overline {\widetilde u}^2 +\overline {\widetilde v}^2 \right)^{ 1/2 } \overline {\widetilde u}, \quad
	\tau_{yz} = - \left( \frac{\kappa}{\ln z_1/z_0} \right)^2 \left( \overline {\widetilde u}^2 + \overline {\widetilde v}^2 \right)^{ 1/2 } \overline {\widetilde v},
	\eeq
	where the overline denotes filtering at a scale of $2\Delta$, $z_1$ is the half vertical grid distance, and $\tau_{xz}$ and $\tau_{yz}$ are the shear stresses in the streamwise and spanwise directions, respectively. The $2\Delta$ filtering was suggested by \citet{bou05} to reduce the log-layer mismatch \cite{bra10}. The reasoning for this procedure is that the filtering preserves the important large scale variations. At the same time, the use of filtered velocities results in an average stress that is very close to the stress predicted by the average similarity formulation for homogeneous
	surfaces \citep{bou05}. The vertical derivative at the first horizontal plane is calculated using the Monin-Obukhov similarity
	theory without stability correction \citep{moe84}, which is implemented following the method proposed by \citet{alb96}.

	\section{Computational setup and simulation results}\label{sec2}

	The computational domain size is $2\pi\, {\rm km} \times 2\pi\, {\rm km} \times 2\, {\rm km}$ in streamwise, spanwise, and vertical directions, respectively. The horizontal domain size is more than six times larger than the boundary layer height to ensure that large horizontal streamwise flow structures are captured appropriately \citep{fos13}. The vertical domain size is more than two times higher than the boundary layer height. The simulations are performed on a $288^3$ grid. The initial potential temperature profile is $\theta(z) = \theta_0 + \Gamma z$, where $\theta_0 = 300$~K is the reference potential temperature and $\Gamma$ is the free-atmosphere lapse rate. The initial velocity profile is $\pu = G \pe_x$, where $G=12$~m/s is the geostrophic wind speed. Small random perturbations are added to the initial fields of $\pu$ and $\theta$ near the surface ($z \le 100$~m) to spin-up turbulence. Statistics are collected over one inertial period, namely $\Delta (ft) = 2 \pi$, after the boundary layer has reached a quasi-stationary state \citep{col92}. 
	A summary of these simulations is given in Table~I of the manuscript.

\begin{figure}[!tb]
		\centering
		\begin{overpic}[width=0.42\textwidth]{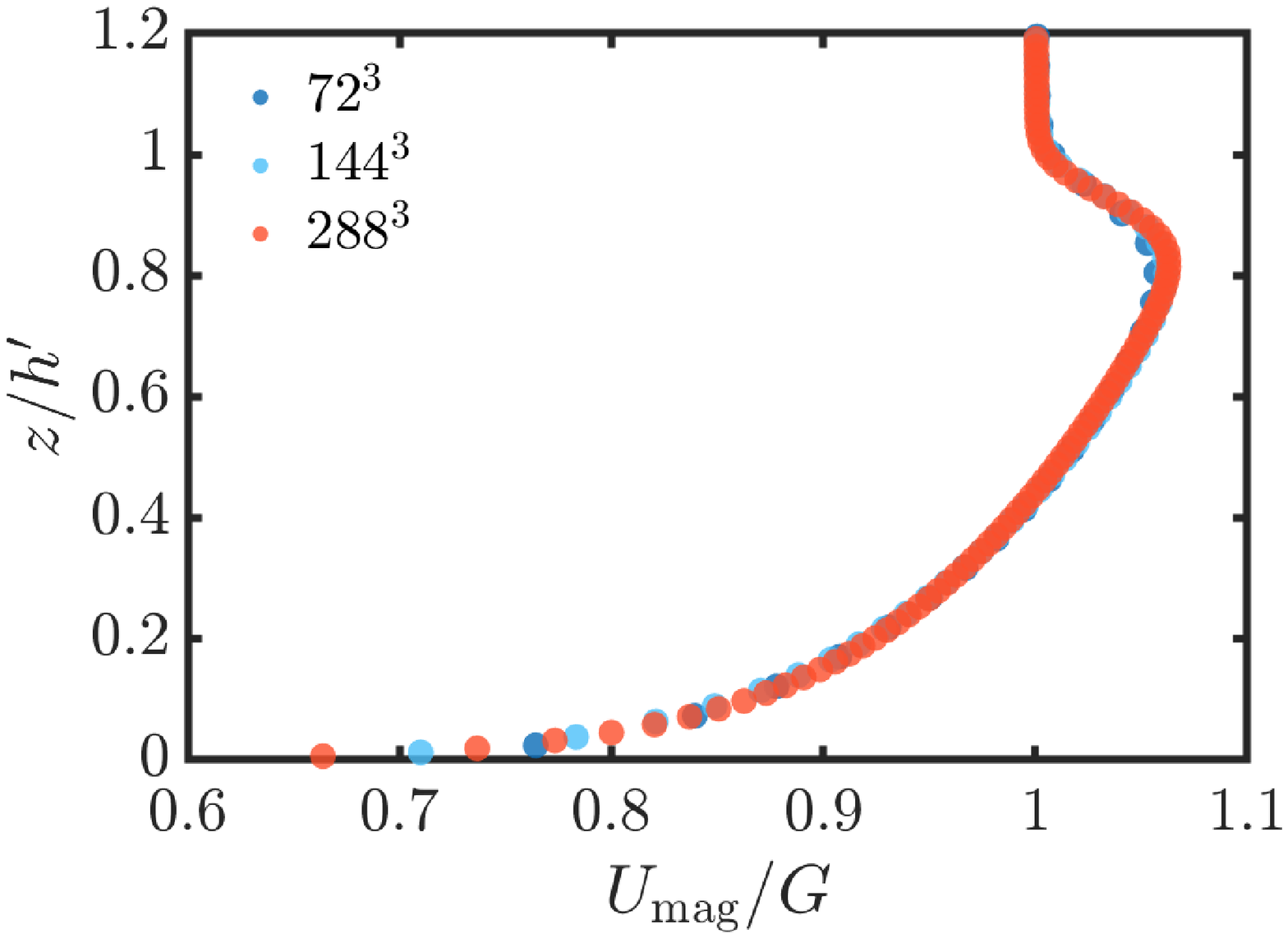} 
			\put(0,66){$(a)$}
		\end{overpic}
		\begin{overpic}[width=0.42\textwidth]{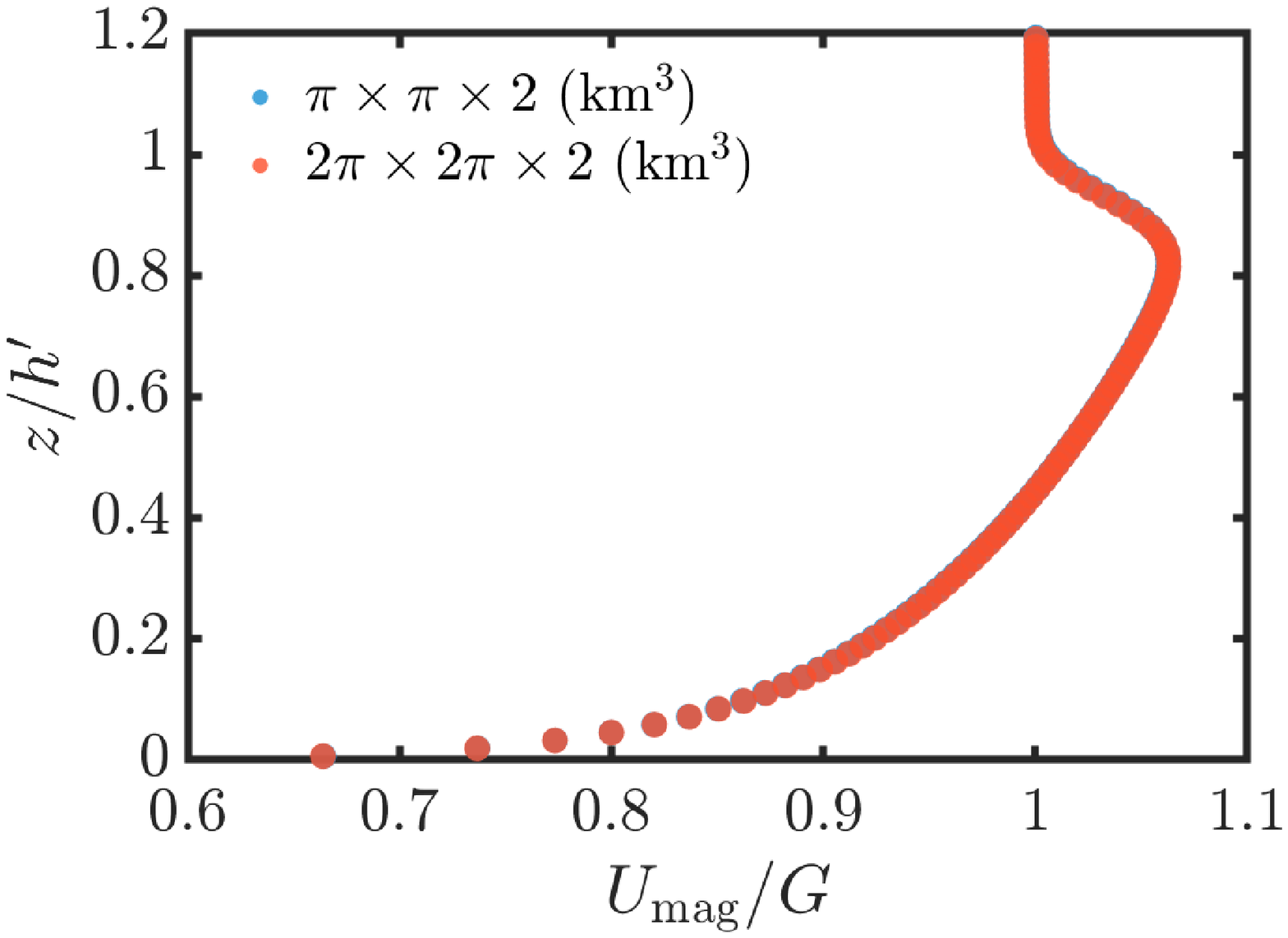} 
			\put(0,66){$(b)$}
		\end{overpic}
		\caption{Vertical profiles of the mean dimensionless wind speed for case E with (a) different grid resolution on a $2\pi \, \textrm{km} \times 2\pi \, \textrm{km} \times 2 \, \textrm{km}$ domain and (b) different computational domain sizes using the same grid spacing $(\Delta x, \Delta y, \Delta z)$. }
		\label{fig.u-grid} 
	\end{figure}

	To confirm that the main findings do not depend on the employed grid resolution and the domain size, we performed simulations for case E (see Table~I of the manuscript) using different grid resolutions ($72^3$, $144^3$, and $288^3$) and for different domain sizes, i.e. $2\pi \, \textrm{km} \times 2\pi \, \textrm{km} \times 2 \, \textrm{km}$ on a $288\times288\times288$ grid and $\pi \, \textrm{km} \times \pi \, \textrm{km} \times 2 \, \textrm{km}$ on a $144\times144\times288$ grid.
	Figure~\ref{fig.u-grid} shows the temporally and horizontally averaged vertical profile obtained from these simulations, which confirms that the obtained velocity profiles do not depend on the employed numerical resolution or domain size.

\begin{figure}[!tb]
		\centering
		\begin{overpic}[width=0.42\textwidth]{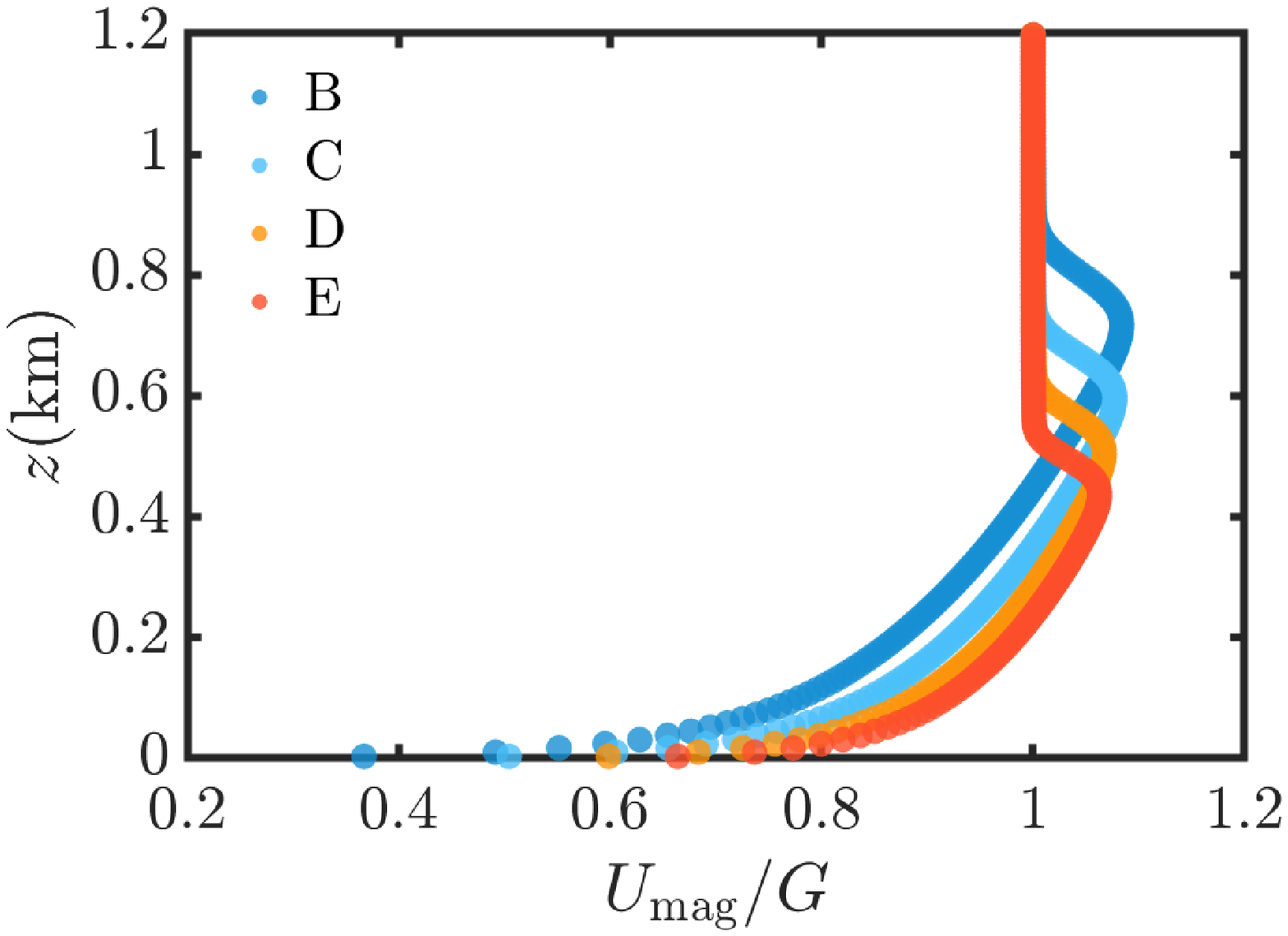} 
			\put(0,66){$(a)$}
		\end{overpic}
		\begin{overpic}[width=0.42\textwidth]{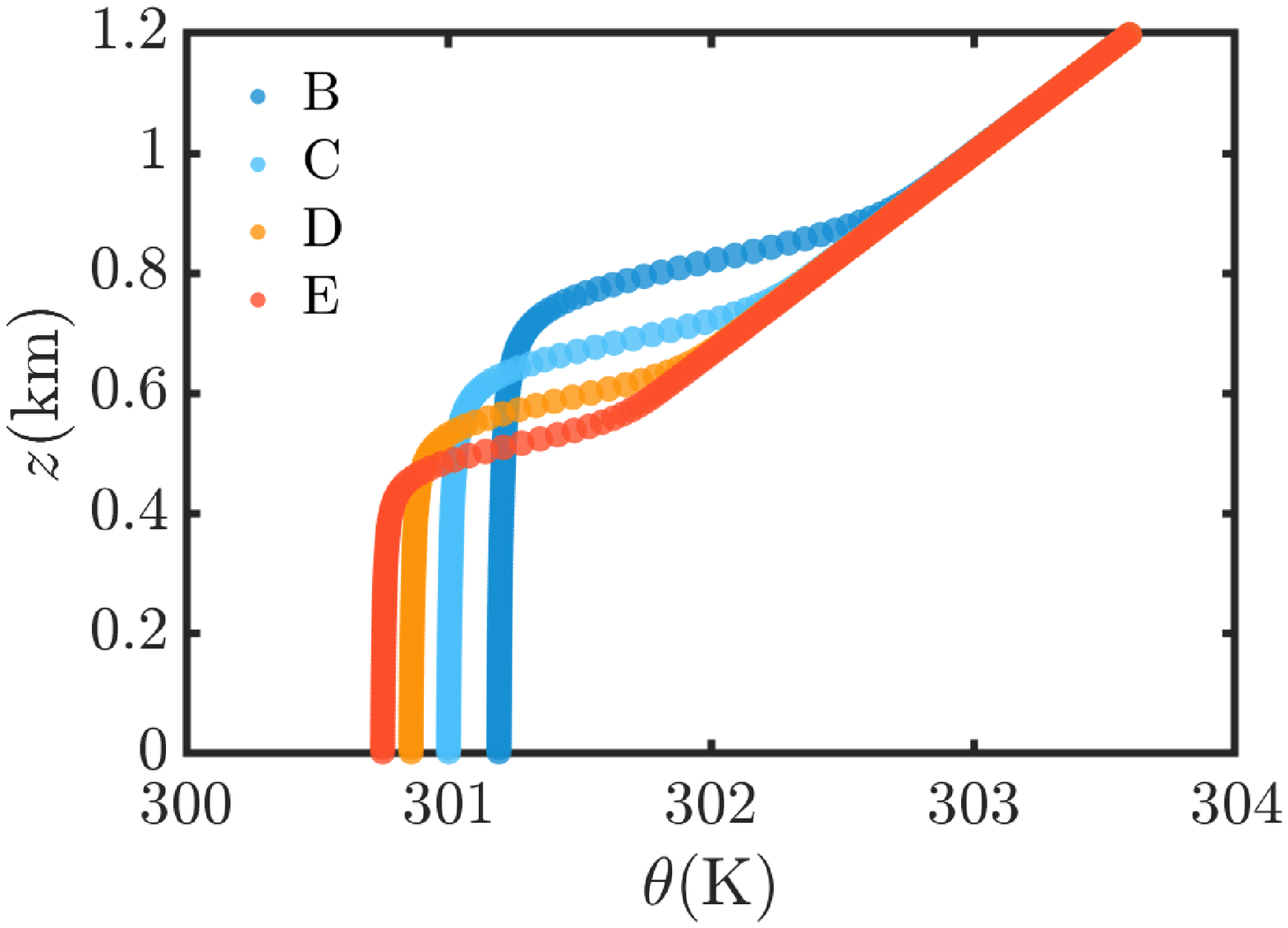} 
			\put(0,66){$(b)$}
		\end{overpic}
		\caption{Vertical profiles of the mean (a) dimensionless wind speed and (b) potential temperature for cases B-E. }
		\label{fig.theta} 
	\end{figure}

	Figure~\ref{fig.theta} shows the vertical profiles of mean wind speed and potential temperature for cases B-E (see Table~I of the manuscript), which have the same atmospheric lapse rate, but different surface roughness lengths.
	The figure shows clearly that the boundary-layer height reduces as the roughness length reduces. Note that a distinct characteristic of the CNBL, as shown in figure~\ref{fig.theta}b, is that the potential temperature is almost uniform below the inversion layer.

	\section{Derivation of the potential temperature flux}\label{sec3}
	
	Below we give a detailed explanation of Eq.~(1) of the manuscript
	and the method we used to determine the values of $s$ and $r$.
	
	For the boundary layers in quasi-equilibrium state, the mean profile is only a function of $z$, which is controlled by the parameters involved in the governing equations and boundary conditions, i.e., 
	\beq
	q = q(z; f, G, \beta, \Gamma, z_0).
	\eeq 
	Therefore the general form of the dimensionless potential temperature can be written as
	\beq\lb{eq.q1}
	{\color{black} \fr{q \beta z_0}{G^3} }  = q_1 \left( \fr{z}{z_0}; Ro_G=\fr{G}{|f| z_0}, Zi=\fr{1}{|f|} \sqrt{ \beta \Gamma } \right).
	\eeq 
	Here and below $q_1$ denotes the functional form of $q$. 
	Using the geostrophic draw law for CNBLs \citep{zil05, liu20d}, 
	\beq 
	\left( \kappa \fr{G}{u_*} \right)^2 = \left[ \ln \left(Ro_G \fr{u_*}{G} \right) - A(Zi) \right]^2 + B(Zi)^2, 
	\eeq  
	where $A=A(Zi)$ and $B=B(Zi)$ are dimensionless coefficients, Eq.~\er{eq.q1} can be written in terms of $u_*$ as 
	\beq\lb{eq.q2}
	{\color{black} \fr{q \beta z_0}{u_*^3} }  = q_1 \left( \fr{z}{z_0}; Ro=\fr{u_*}{|f| z_0}, Zi \right).
	\eeq 
	In addition, the boundary-layer height for CNBLs can be estimated as \citep{zil07}
	\beq \lb{eq.h'}
	\left( Ro \fr{z_0}{h'} \right)^2 = \fr{1}{C_R^2} + \fr{1}{C_N^2} Zi,
	\eeq 
	where $C_R$ and $C_N$ are empirical constants. 
	Thus, Eq.~\er{eq.q2} can be written in terms of $h'$ as follows 
	\beq\lb{eq.q3}
	{\color{black} \fr{q \beta z_0}{u_*^3} }  = q_1 \left( \xi = \fr{z}{h'}; Ro, Zi \right) = -\Pi(\xi; Ro, Zi).
	\eeq 
	
	We assume that there is a power law dependence between $\Pi$ and the two control parameters $Ro$ and $Zi$,
	\beq\lb{eq.q4}
	{\color{black} \fr{q \beta z_0}{u_*^3} }   = -\Pi(\xi; Ro, Zi) = -Ro^r Zi^s \Pi_1(\xi).
	\eeq 
	To determine the values of the power indices $r$ and $s$ in Eq.~\er{eq.q4}, we take the derivative of Eq.~\er{eq.q4}, which gives 
	\beq\lb{eq.q5}
	\ln \left( - {\color{black} \fr{q' \beta z_0}{u_*^3} }   \right) = \ln c_\Pi + r \ln Ro + s \ln Zi, \quad 
	q'=\fr{ \textrm{d} q}{ \textrm{d} \xi}.
	\eeq 
	This indicates that $r$ and $s$ can be determined from the slope of the curve $\color{black} \ln ( - q' \beta z_0 / u_*^3 )$ vs $\ln Ro$ and $\color{black} \ln ( - q' \beta z_0 / u_*^3 )$ vs $\ln Zi$, respectively, in the surface layer. Using a least-squares fitting, we find that (see figure 2 of the manuscript)
	\beq 
	{\color{black} r=-1.002 \approx -1,} \quad 
	s=1.004 \approx 1.
	\eeq 
	Note that the value of $c_\Pi$ can also be determined from Eq.~\er{eq.q5} once $r$ and $s$ have been determined. The corresponding results are shown in Table~I of the manuscript.

	\section{The dependence of small parameter on the control parameter}\label{sec4}
	
	We remark that $\epsilon$ in Eq.~(2) of the manuscript depends only on the control parameter $Zi$. To prove it, we recall that the boundary-layer height for CNBLs can also be defined as the height at which the potential temperature flux reaches its minimum value. We denote this height as $h_t$, which can be estimated as \citep{zil07, abk13}
	\beq \lb{eq.ht}
	\left( Ro \fr{z_0}{h_t} \right)^2 = \fr{1}{C_R'^2} + \fr{1}{C_N'^2} Zi,
	\eeq 
	where $C_R'$ and $C_N'$ are empirical constants. Combining equations~\er{eq.h'} and \er{eq.ht} gives
	\beq 
	2 \epsilon \equiv 1-\fr{h_t}{h'} = 1 - \left( \fr{ \fr{1}{C_R^2} + \fr{1}{C_N^2} Zi }{ \fr{1}{C_R'^2} + \fr{1}{C_N'^2} Zi } \right)^{1/2}.
	\eeq 
	Note that the variance of $\epsilon$ with $Zi$ is very small (see Table~I of the manuscript) and therefore we take it as a constant in this work.

	\section{Adopted formulations of the existing wind speed models}\label{sec5}
	
	Here we describe the previous models for the wind speed profiles in CNBLs, i.e. \citet{zil05}, \citet{gry07}, and \citet{kel19}, which are shown in figure 4 of the manuscript.

	\subsection{Zilitinkevich and Esau (Z2005 in figure 4 of the manuscript) }
	
	\citet{zil05} showed that the wind shear in the surface layers can be written as
	\beq \lb{eq.zil-1}
	\fr{\textrm{d}  U_{\rm mag}}{\textrm{d} z} = \fr{\sqrt \tau}{\kappa} \left( \fr{1}{z} + \fr{C_u}{L_M} \right),
	\eeq 
	where $C_u=2.5$ is an empirical constant and $L_{M}$ is the combined turbulent length scale. In CNBLs, 
	\beq 
	\fr{L_f^2}{L_M^2} = 1 + \left(  C_m Zi \right)^2, \quad 
	L_f = \fr{u_*}{|f|},
	\eeq 
	where $C_m=0.1$ is an empirical constant. Following \citet{gry07}, $\sqrt \tau$ is approximated as
	\beq\lb{eq.tau1}
	\sqrt \tau = u_* \left( 1- \fr{z}{h} \right),
	\eeq 
	where $h$ is the boundary layer height at which the total momentum flux reaches 5\% of its surface value. 
	By substituting Eq.~\er{eq.tau1} into Eq.~\er{eq.zil-1}, and after vertical integration we obtain
	\beq \lb{eq.zil-2}
	\fr{\kappa U_{\rm mag} }{u_*} = \ln \left( \fr{z}{z_0} \right) + \left( \fr{C_u}{L_M} - \fr{1}{h} \right) z - \fr{C_u}{2 L_M h} z^2.
	\eeq

	\subsection{Gryning et al. (G2007 in figure 4 of the manuscript) }
	
	\citet{gry07} showed that the wind shear in the surface layer can be written as 
	\beq \lb{eq.gry-1}
	\fr{\textrm{d}  U_{\rm mag}}{\textrm{d} z} = \fr{\sqrt \tau}{\kappa} \fr{1}{l},
	\eeq 
	where $l$ is the local length scale that can be modelled by a combination of the length scale in the surface layer $L_{\rm SL}$, in the middle of the boundary layer $L_{\rm MBL}$, and in the upper part of the boundary layer $L_{\rm UBL}$, i.e.
	\beq\lb{eq.l}
	\fr{1}{l} = \fr{1}{L_{\rm SL}} + \fr{1}{ L_{\rm MBL} } + \fr{1}{ L_{\rm UBL} }.
	\eeq 
	In CNBLs, 
	\beq\lb{eq.l2}
	L_{\rm SL} = z, \quad 
	L_{\rm UBL} = h-z.
	\eeq 
	By substituting Eqs.~\er{eq.tau1}, \er{eq.l} and \er{eq.l2} into Eq.~\er{eq.gry-1} and after vertical integration we obtain
	\beq \lb{eq.gry-2}
	\fr{\kappa U_{\rm mag} }{u_*} = \ln \left( \fr{z}{z_0} \right) + \fr{z}{ L_{\rm MBL} } \left( 1 - \fr{z}{2h} \right).
	\eeq 
	The length scale $L_{\rm MBL}$ is parameterized, for details see \citet{gry07} and \citet{ped14}, such that the wind speed at the top of the boundary layer conforms to the geostrophic wind speed.

	\subsection{Kelly et al. (K2019 in figure 4 of the manuscript) }
	
	\citet{kel19} proposed the wind speed profile as follows 
	\beq \lb{eq.kel}
	\fr{\kappa U_{\rm mag} }{u_*} = \ln \left( \fr{z}{z_0} \right) + \fr{a}{2}  \left( \fr{z}{ L_{\rm TD} } \right)^2,
	\eeq 
	where $a=4.3$ and 
	\beq 
	L_{\rm TD}^2 = \fr{L_f^2}{\Pi_\theta Zi^2 Ro_{h_\theta}^{0.3} }, \quad 
	L_f = \fr{u_*}{|f|}.
	\eeq 
	Here $\Pi_\theta =-0.0016$ and $Ro_{h_\theta} = L_f / h_\theta$ is the Rossby number based on the boundary-layer height $h_\theta$, which is defined by the height at which the gradient of the potential temperature reaches its maximum value. 
	
	\bibliography{literature_windfarms}